\documentclass[12pt]{article}
\usepackage{amsmath,amssymb}
\usepackage{graphicx,color}
\numberwithin{equation}{section}
\usepackage{cite}
\usepackage{bm}
\usepackage{dcolumn}  
\newcommand{\be}{\begin{equation}}
\newcommand{\ee}{\end{equation}}
\newcommand{\bea}{\begin{eqnarray}}
\newcommand{\eea}{\end{eqnarray}}
\newcommand{\beqar}{\begin{eqnarray*}}
\newcommand{\eeqar}{\end{eqnarray*}}

\begin{document}
\baselineskip 18pt%
\begin{titlepage}
\vspace*{1mm}%
\hfill%
\vspace*{15mm}%
\hfill
\vbox{
    \halign{#\hfil         \cr
          } 
      }  
\vspace*{20mm}

\centerline{{\Large {\bf Toroidal solutions in Ho\v{r}ava Gravity }}}
\centerline{{\Large{\bf    }}}
\vspace*{5mm}
\begin{center}
{ Ahmad Ghodsi}\\
{ a-ghodsi@ferdowsi.um.ac.ir}\\
\vspace*{0.2cm}
{ Department of Physics, Ferdowsi University of Mashhad, \\
P.O. Box 1436, Mashhad, Iran}\\
\vspace*{0.1cm}
\end{center}

\begin{abstract} 
Recently a new four-dimensional non relativistic renormalizable theory of gravity was proposed by Horava. This gravity reduces to Einstein gravity at large distances. In this paper by using the new action for gravity we present different toroidal solutions to the equations of motion. Our solutions describe the near horizon geometry with slow rotating parameter. 
\end{abstract} 

\end{titlepage}

\section{Introduction}
Recently a new four-dimensional non relativistic renormalizable theory of gravity  was proposed by
Ho\v{r}ava \cite{Horava:2009uw}. It is believed that this theory is a UV completion for the Einstein
theory of gravitation. Recently a lot of efforts have been done to understand this theory, \cite{Lu:2009em,Horava:2008ih,Horava:2009if,Takahashi:2009wc, Calcagni:2009ar,Kiritsis:2009sh,
Kluson:2009sm,Mukohyama:2009gg,Nikolic:2009hr,Brandenberger:2009yt,Nikolic:2009jg
,Izawa:2009ne,Nastase:2009nk,Pal:2009np,Cai:2009pe,Cai:2009ar,Volovik:2009av,Piao:2009ax,
Gao:2009bx,Colgain:2009fe,Sotiriou:2009gy,Chen:2009ka,Mukohyama:2009zs,Myung:2009dc,Cai:2009dx,
Orlando:2009en,Gao:2009er,Ha:2009ft,Nishioka:2009iq,Kehagias:2009is,Konoplya:2009ig,Ghodsi:2009zi}. 
In \cite{Lu:2009em} the solutions with spherical symmetry has been found. It also presents equations of motion for Horava gravity. The topological black hole solutions has been found in \cite{Cai:2009pe}. In this paper, in section two, we review briefly the static toroidal solution. This is a special solution found in \cite{Cai:2009pe}. In section three we try to find the rotational solutions. We use the equations of motion presented in \cite{Lu:2009em} and show that there are different possible solutions to the equations of motion.  

We start from the four-dimensional metric written in the ADM formalism, \cite{Arnowitt:1962hi}
\be
ds_4^2= - N^2  dt^2 + g_{ij} (dx^i - N^i dt) (dx^j - N^j dt)\,.
\ee
The Einstein-Hilbert action in this formalism is given by
\be
S_{EH} = \frac{1}{16\pi G} \int d^4x \sqrt{g} N (K_{ij} K^{ij} - K^2 + R - 2\Lambda)\,,
\ee
where $G$ is the four dimensional Newton's constant and $K_{ij}$ is the second fundamental form and is defined by
\be
K_{ij} = \frac{1}{2N} (\partial_t g_{ij} - \nabla_i N_j - \nabla_j N_i)\,.
\ee
The action proposed by Ho\v{r}ava is a  non-relativistic renormalizable gravitational theory and is given by \cite{Horava:2009uw}
\bea
S&=&\int dtd^3x\sqrt{g}N\left\{\frac{2}{\kappa^2}(K_{ij}K^{ij}
-\lambda K^2)+\frac{\kappa^2\mu^2(\Lambda_W R
-3\Lambda_W^2)}{8(1-3\lambda)}+\frac{\kappa^2\mu^2
(1-4\lambda)}{32(1-3\lambda)}R^2\right.\,\nonumber\\
&&\left.\quad\quad\quad\quad\quad\quad-\frac{\kappa^2\mu^2}{8}R_{ij}R^{ij}
+\frac{\kappa^2\mu}{2w^2}\epsilon^{ijk}R_{i\ell}\nabla_jR_k^\ell
-\frac{\kappa^2}{2w^4}C_{ij}C^{ij}\right\}\,,\label{act}
\eea
where $\lambda\,,\kappa\,,\mu\,,w$ and $\Lambda_W$ are constant
parameters, and $C_{ij}$ is the Cotton tensor, defined by
\be
C^{ij}=\epsilon^{ik\ell}\nabla_k\left(R^j{}_\ell
-\frac14R\delta_\ell^j\right)=\epsilon^{ik\ell}\nabla_k R^j{}_\ell
-\frac14\epsilon^{ikj}\partial_kR\,.
\ee
Using the relation 
\be
\epsilon^{ijk}R_{i\ell}\nabla_jR^\ell{}_k=R_{i\ell}\left[C^{i\ell}
-\frac14\epsilon^{ij\ell}\partial_j R\right]=R_{i\ell}C^{i\ell}\,,\ee
one can rewrite the action (\ref{act}) as
\bea
S&=&\int dtd^3x\, ({\cal L}_0 + {\cal L}_1)\,,\nonumber\\ {\cal L}_0
&=& \sqrt{g}N\left\{\frac{2}{\kappa^2}(K_{ij}K^{ij} -\lambda
K^2)+\frac{\kappa^2\mu^2(\Lambda_W R
  -3\Lambda_W^2)}{8(1-3\lambda)}\right\}\,,\nonumber\\ {\cal L}_1&=&
\sqrt{g}N\left\{\frac{\kappa^2\mu^2 (1-4\lambda)}{32(1-3\lambda)}R^2
-\frac{\kappa^2}{2w^4} \left(C_{ij} -\frac{\mu w^2}{2}R_{ij}\right)
\left(C^{ij} -\frac{\mu w^2}{2}R^{ij}\right)\right\}\,.\label{action}
\eea%
By comparing ${\cal L}_0$ with the general theory of relativity in the ADM formalism, one can read the speed of light, the Newton's constant and the cosmological constant as
\be
c=\frac{\kappa^2\mu}{4} \sqrt{\frac{\Lambda_W}{1-3\lambda}}\,,\qquad
G=\frac{\kappa^2}{32\pi\,c}\,,\qquad
\Lambda=\frac32 \Lambda_W\,.
\ee
Additionally, demanding that ${\cal L}_0$ gives the usual four dimensional Einstein-Hilbert Lagrangian (general covariance), one finds that $\lambda=1$.
\section{Toroidal static solution}
The topological black hole solution has been found in \cite{Cai:2009pe}. We are interested to the special case of toroidal symmetric solutions in this paper. So in this section we review the special solution found in \cite{Cai:2009pe} with torodial symmetry. Starting from the following ansatz
\be
ds^2=-N(r)^2 dt^2+\frac{dr^2}{f(r)}+r^2({d\theta^2+d\phi^2})\,,
\ee
and insert it into the total Lagrangian ${\cal L}_0+{\cal L}_1$. Due to the special form of the ansatz, the Cotton tensor is zero.
The Lagrangian with general value for $\lambda$ is given by
\bea
{\cal L}_0+{\cal L}_1&=&\frac{3\mu^2\kappa^2 N}{8(-1+3\lambda)r^2 f^\frac12}\bigg(\frac{(1-\lambda)r^2}{6}{f'}^2 +\frac23 r (\Lambda_W r^2+\lambda f)f'+\frac13(1-2\lambda)f^2\nonumber\\
&+&\frac23\Lambda_W r^2 f+\Lambda_W^2 r^4\bigg)\,,
\eea
where prim denotes the derivative with respect to $r$. The solution to the equations of motion is \cite{Cai:2009pe}
\bea
f(r)&=&-M r^n-\Lambda_W r^2\,,\quad
N^2(r)=f(r) (Cr)^{2(1-2n)}\,,\quad
n=\frac{2\lambda-\sqrt{-2+6\lambda}}{-1+\lambda}\,,\label{soll}
\eea
where, $M$ and $C$ are the constants of integrations.

The above solution has two real roots for $M>0$ and $\Lambda_W<0$ at $r_{-}=0$ and $r_+=(-\frac{M}{\Lambda_W})^{\frac{1}{2-n}}$. The scalar curvature is given by ${\cal{R}}=2(3\Lambda_W+M(n+1)r^{{n-2}})$ and because  $\lambda\rightarrow +\infty$ then $n\rightarrow 2$ so we always have a curvature singularity at $r=0$. When $\Lambda_W>0$ then $r=0$ is a naked singularity. 

\section{Rotating solutions}
After a brief review of the toroidal static solution, in this section we try to find other solutions to the Horava gravity by including the rotation. Because of the rotation, we have not enough symmetry to apply the previous method (i.e. insert the ansatz into the Lagrangian), instead we must solve the equations of motion directly. The equations of motion are very difficult to solve since they are up to six derivatives and the metric in the rotating solutions depend to the rotation coordinate as well as the radial coordinate. To overcome this difficulty we try to find the near horizon geometry of the rotating black holes in this paper. This will simplify the equations of motion since as we will see in what follows, the functional form of the solutions with respect to the radial coordinate will be fixed, so it remains to find their dependence on the rotation coordinate.
\subsection{Extremality}
To find the radial behavior of the extremal solutions we start to find the extremality condition for the general solution found in (\ref{soll}). We first find the temperature of the solution (\ref{soll}). The temperature of this black hole can be computed by finding the surface gravity at the horizon. The result will be
\bea
T&=&\frac{1}{2\pi}\left(2\Lambda_W (n-1)r_0^{-2n+2}+(\frac32 n-1)M r_0^{-n}\right)\cr &&\cr
&=&\frac{\Lambda_W(\frac{n}{2}-1)}{2\pi}\bigg(-\frac{M}{\Lambda_W}\bigg)^{\frac{2(n-1)}{n-2}}\,,
\eea
where the last equality coming from the fact that the location of the horizon is at $r_0=r_+$. The extremality condition happens when the temperature is zero, so we find the critical value of $M$ for an extremal solution to be zero. 
In this way the geometry of the extremal solution will be
\be
ds^2=r^{4(1-n)}dt^2-\frac{dr^2}{\Lambda_W r^2}+r^2(d\theta^2+d\phi^2)\,.
\ee
\subsection{Two derivative solutions}
Before we start to solve the equations of motion, we consider the special case where the equations of motion just contain up to two derivative terms. In this case we expect that we find the known solutions for the Einstein gravity. 
The solution to the equations of motion for the Einstein gravity is given by \cite{Caldarelli:1998hg}
\be
ds^2=-N^2 dt^2+\frac{\rho^2}{\Delta_r} dr^2+\frac{\rho^2}{\Delta_\theta} d\theta^2+\frac{\Sigma^2}{\rho^2}(d\phi-\varpi dt)^2\,,
\ee
with
\bea
\rho^2&=&r^2+a^2 \theta^2\,, \quad \Delta_\theta=1+\frac{a^2}{l^2}\theta^4\,,\quad \Delta_r=a^2-2 M r+\frac{r^4}{l^2}\,, \nonumber\\ \Sigma^2&=&r^4\Delta_\theta-a^2 \theta^4 \Delta_r\,,\quad \varpi=\frac{\Delta_r \theta^2+r^2\Delta_\theta}{\Sigma^2}a\,,\quad N^2=\frac{\rho^2 \Delta_\theta \Delta_r}{\Sigma^2}\,,
\eea
where $a$ is the rotation parameter and in our notation $l^2=-\frac{2}{\Lambda_W}$. We are interested to find the extremal solution and its near horizon geometry. The extremal condition happens
when 
\be
M=\frac12\frac{a^2 l^2+ r_0^4}{r_0l^2}\,,\quad r_0^2=\frac{1}{\sqrt{3}} a l\,,
\ee
where $r_0$ is the location of the horizon. For finding the near horizon geometry we need to change our variables to some new dimensionless coordinates as follows
\be
r=r_0+\frac{\epsilon}{y} a\,,\quad t=\frac{c_0}{\epsilon}\tau\,,\quad\phi=\hat{\phi}+\frac{\sqrt{3}c_0}{l\epsilon} \tau\,,\quad c_0^2=\frac{r_0^2}{12}\,.
\ee
Sending $\epsilon\rightarrow 0$ one finds the following metric
\be
ds^2=(1+\frac{a^2\theta^2}{r_0^2})\bigg(-\frac{1}{2\sqrt{3}}\frac{a^3}{l y^2} d\tau^2+\frac{l^2}{6 y^2} dy^2+\frac{r_0^2}{1+\frac{a^2}{l^2}\theta^4} d\theta^2+r_0^2\frac{1+\frac{a^2}{l^2}\theta^4}{(1+\frac{a^2}{r_0^2}\theta^2)^2}(d\hat{\phi}+\frac{a}{ly}d\tau)^2\bigg)\,.\label{exrot}
\ee
This is the near horizon geometry of the rotating black holes with toroidal symmetry. We expect that this satisfies the equations of motion up to two derivative terms. As a double check, we have inserted this solution to the equations of motion and they satisfy these equations.

\subsection{Higher derivative solutions}
We are interested to find the effect of higher curvature terms. To find the solutions we use the following steps

1. In the rotating solutions the Cotton tensor is not necessarily zero and this makes the problem difficult to solve. To find the rotating solution, we consider the slow rotation condition, i.e. $a\ll l$ as a parameter of perturbation and solve the equations of motion up to ${\mathcal{O}}(a)$.

2. For finding the extremal rotating solution we use the tree-level solution (\ref{exrot}) as our guide. We start from  the following ansatz
\be
ds^2=-\frac{A_1^2(\theta)}{y^2}d\tau^2+\frac{A_2(\theta)}{y^2} dy^2+A_3(\theta)d\theta^2+ A_4(\theta)(d\hat{\phi}+\frac{a}{l y}d\tau)^2\,,
\ee 
where $y$ is the radial near horizon coordinates and the other functions in the metric are some general functions.
This metric satisfies the equation of motion coming from variation of the Lagrangian with respect to $N$, the laps function. So we just need to insert this general ansatz into the other equations of motion coming from variation with respect to the shift functions $N^i$ and the metric $g^{ij}$.

3. One may notice that we have a freedom for time scaling in the metric. We have fixed this by choosing the above proper off-diagonal term. 

4. There is another freedom when one chooses the function $A_3(\theta)$. Because this is just a field redefinition, all different functions of $\theta$ will be equivalent by a change of coordinate on $\theta$. For fixing this freedom  we suppose the following functional form 
\be
A_3(\theta)=r_0^2\frac{1+\frac{a^2}{r_0^2}\theta^2}{1+\frac{a^2}{l^2}\theta^4}\,,
\ee
where we have chosen it in such a way that we can compare the new metric with the previous two derivative case. 

5. For solving the equations of motion perturbativly in terms of the rotating parameter $a$, we choose polynomial functions with unknown constant coefficients as 
\be
A^2_1(\theta)=s_1 a^3 (1+b_1 a \theta^2)\,,\quad A_2(\theta)=s_2 (1+b_2 a \theta^2)\,,\quad A_4(\theta)=s_4 a (1+b_4 a \theta^2)\,.
\ee
where in writing these functions we have used the fact that we have a symmetry under $(\theta\leftrightarrow-\theta)$. 

6. Similar to (\ref{exrot}) the regularity condition implies (see e.g. \cite{Astefanesei:2006dd} ) 
\be
A_1(\theta)A_2^\frac12(\theta)\rightarrow constant\,,\quad
\frac{A_3(\theta)}{A_4(\theta)}\rightarrow1\,,\qquad \theta\rightarrow 0\,,
\ee
which gives a simple constraint as $s_4=\frac{r_0^2}{a}$. 

7. Similar to the Einstein gravity solution, we suppose that $r_0^2=z a$ with $z$ as a function of the constants of the Horava gravity. In fact this is nothing but the normalization for $A_3(\theta)$.

By considering all these facts, we find the four Algebraic equations (see appendix A). As one sees there are four equations and $s_1, s_2, b_1, b_2, b_4 $ and $s_4=z$ as unknow constants. Because already we have used all symmetries and boundary conditions there are no more constraints left. 

\subsubsection{$w$ independent solution}

One amazing observation of the equations shows that when $b_2=b_4$ then the constants are independent of $w$. At this step even before solving the equations of motion one can verify that the Cotton tensor is zero for this ansatz. To find the solution to the equations of motion we do the following steps. We begin by solving the first three equations and find the following values for $s_1, b_1$ and $b_2$ 
\bea
\!\!\!\!&&b_1=\frac{-\frac23(l^2-6s_2)}{(\lambda-\frac13)l^2 z s_2((\lambda-1)l^2+4s_2)}((\lambda-1)^2l^4-\frac32(\lambda-1)(z^2-\frac83s_2)l^2+12(\lambda-\frac56)z^2s_2)\,,\cr\nonumber
\!\!\!\!&&b_2=-\frac{2z(l^2-6s_2)}{l^2((\lambda-1)l^2+4s_2)}\,,\qquad
s_1=\frac18\frac{(l^2(\lambda-1)+4s_2)^2z}{((\lambda-1)l^4+8l^2s_2-24s_2^2)s_2}\,,
\eea
and then we insert these values into the fourth equation, which gives the following equation for $s_2$
\be
(\lambda-1)l^6+(-8\lambda+12)s_2l^4-48l^2s_2^2+96s_2^3=0\,,
\ee
whereit is independent of $z$. The above solution is a family of one parameter solutions (just depend on $z$) with zero Cotton tensor. In the special case of $\lambda=1$ the above solution will simplifies to
\be
b_1=b_2=\frac{z}{l^2}\frac{1\pm\sqrt{3}}{1\pm\frac{\sqrt{3}}{3}}\,,\quad s_1=\frac{z}{l^2(1\mp\sqrt{3})}\,,s_2=\frac14(1\pm\frac{\sqrt{3}}{3})l^2\,.\label{bb}
\ee

As a special point on this family of the solutions and as an example one may choose $b_1=b_2=b_4$. Again if we solve the equations of motion we will find the following values for a general value of $\lambda$ after solving the first three equations in the Appendix
\bea
&&b_1=b_2=b_4=-\frac{4}{3z}\,,\cr\nonumber &&s_1=\frac{9}{64}\frac{(-1+3\lambda)z^5(4l^2+9z^2)}{l^2((\lambda-1)l^2-\frac32z^2)((\lambda-1)l^4-3l^2z^2-\frac{27}{8}z^4)}\,\cr\nonumber
&&s_2=-\frac12\frac{l^2(2l^2(\lambda-1)-3z^2)}{4l^2+9z^2}\,.
\eea 
Inserting the above values into the fourth equation gives an equation for $z$
\be
(\lambda-1)^2 l^6 -6 (\lambda-\frac34)l^4 z^2-\frac{27}{4}(\lambda-1)l^2 z^4+\frac{81}{32} z^{6}=0\,.
\ee
This equation shows that the location of the horizon depends on $\lambda$ and $l$. Again the special case $\lambda=1$ gives the values found in (\ref{bb}) with $z=\pm\frac{2}{3^{\frac34}}\sqrt{\pm l^2}$.

\subsubsection{$w$ dependent solution}
In general, when one chooses $b_2\neq b_4$, the constant values will be $w$-dependent. In this case one may solve the equations of motion and find the first three equations for $s_1,s_2$ and $b_1$ in terms of $b_2,b_4$ and $z$. Putting them into the fourth equation gives a relation between the remaining free parameters. This an equation of degree 8 for $z$, 6 for $b_2$ and 5 for $b_4$, so impossible to solve!. 

To find a solution we restrict ourselves to a special limit of parameters. One possible solution could be found as a series of $\frac{1}{w^4}$. Also we consider the location of the horizon $r_0$, to be independent of $w$ and its value is the same as $w$-independent solution. With these simplifications we find  the following solution to the order of $\mathcal{O}(\frac{1}{w^4})$, in the case of $\lambda=1$,
\bea
b_1&=&-\frac{2}{3^\frac14 l}(1+\frac{x_1}{w^4})\,,\quad b_2=-\frac{2}{3^\frac14 l}(1+\frac{x_2}{w^4})\,,\quad b_4=-\frac{2}{3^\frac14 l}(1+\frac{x_4}{w^4})\nonumber\\
s_1&=&\frac{2}{3^\frac14 l (3+\sqrt{3})}(1+\frac{y_1}{w^4})\,,\quad s_2=\frac{l^2(3+\sqrt{3})}{12(2+\sqrt{3})}(1+\frac{y_2}{w^4})\,,
\eea
with
\bea
x_1&=&-\frac{1}{13} (105\sqrt{3}+217)y_2\,,\quad x_2 = -\frac{1}{13}(45\sqrt{3}+67)y_2\,,\nonumber\\
\quad x_4&=& -(5\sqrt{3}+7)y_2\,,\,\,\qquad\qquad y_1 = \frac{1}{13}(62\sqrt{3}-27)y_2\,,
\eea
where the constant $y_2$ although is arbitrary but can be absorbed into $w$ by a rescaling, so we can set it to one. As we see this will produce the $w$-independent solution when one sends $w$ to the infinity.
\section{Conclusion}
In this paper we have studied the toroidal solutions for the non relativistic and renormalizable theory of gravity proposed by Horava \cite{Horava:2009uw}. We solved equations of motion by using an ansatz with toroidal symmetry. We show our results for general parameters in the theory and in ``detailed balance". 

The static case found in \cite{Cai:2009pe} shows the existence of black hole solutions where their location of horizon depends on the parameters of the theory, when $\Lambda_W<0$. It shows that for $\Lambda_W>0$ we have naked singularities.

In this paper we find the near horizon geometry of the rotating black hole solutions with small rotating parameter $a$ with respect to $l=\sqrt{-\frac{2}{\Lambda_W}}$. So our solution is a series solution in terms of $a$. Also we have assume the $\theta$ to $-\theta$ symmetry. By imposing these constraints we have found a set of Algebraic equations of motion. There is an interesting observation in our solutions for equations of motion. There are two types of solutions. The first one is independent of $w$ parameter and the Cotton tensor for this solution is zero. The other solution depends on $w$ and at $w\rightarrow \infty$ this solution backs to the first solution.



Comparing these results with those found in the two derivative case, one observes that the location of the horizon is shifted due to the higher derivative corrections. This is in agreement with the results found for the spherical solutions  in \cite{Lu:2009em}.

It will be interesting to find the exact rotating solution. In this case the metric will be a complicated function of $\theta$ and $y$. In finding our solutions we have supposed several assumptions, $a\ll l$, $\theta$ to $-\theta$ symmetry and in the case of $w$ dependent solution, the location of the horizon is considered to be independent of $w$. In general there is no reason to have these constraints in the exact solution . So the near horizon of the exact solution just with the above assumptions must agree with our solutions. Note that the regularity condition and the field redefinitions for $t$ and $\theta$ must hold in the exact solution.

\section*{Acknowledgment}
This work was supported by the grant (P/962, 88/12/08) from Ferdowsi University of Mashhad.

\section*{Appendix A} 
\bea
& &\bigg\{(({{b_2}}^{2}-{b_2}{b_4}+{{b_4}}^{2} ) \lambda-\frac12{{b_4}}^{2}-\frac12{{b_2}}^{2}) {l}^{4}+2{z} ( {b_2}+{b_4} ) {l}^{2}-6{{z}}^{2} \bigg\}{s_1} {s_2}+\frac14 ( -1+3\lambda ) {{z}}^{3}=0\,,\nonumber
\eea
\bea
\!\!\!& &\!\!\!{l}^{2}{\kappa}^{4}{s_1} {{s_2}}\bigg\{-{\frac{19}{3}}+({b_1}-5{b_2}-\frac{16}{3}{b_4}){z}+(-\frac{14}{3}{{b_2}}^{2}+({b_1}
-\frac{11}{3}{b_4}){b_2}+({b_1}-5{b_4}){b_4}){{z}}^{2}\bigg\}\nonumber\\
\!\!\!&+&\!\!\!\frac{4}{3} \bigg\{ \frac18 (3\lambda-1) {z}^{4} +\bigg[ -3{{z}}^{3}+{l}^{2}({b_1}+ {b_4}){{z}}^{2}+ \bigg(  [ -\frac{9}{2}{{b_2}}^{2}+ ( {b_1}+\frac12{b_4}) {b_2}-\frac12 ({b_1} -5{b_4} ){b_4}  ] \lambda\nonumber\\
\!\!\!&+&\!\!\!\frac94{{b_2}}^{2}-\frac14{{b_4}}^{2}-\frac12({b_1}+{b_4}){b_2} \bigg){l}^{4} {z}-4{l}^{4}\bigg(  ( {b_2}-\frac12{b_4} ) \lambda-\frac12{b_2}\bigg) \bigg]{s_1} {s_2}\bigg\} \frac{{{z}}^{2}{w}^{4}}{(3\lambda-1) ( {b_2}-{b_4} )}=0,\nonumber
\eea
\bea
\!\!\!& &\!\!\!{l}^{2}{\kappa}^{4}{s_1}(-4{z}+{s_2}({b_2}-{b_4}))(1+({b_2}+{b_4}){z})
+2\bigg\{\bigg[\bigg([{b_2}-(2{b_2}-{b_4})\lambda]{z}-[\frac12({{b_2}}^{2}+{{b_2}}^{2})\lambda\nonumber\\
\!\!\!&-&\!\!\!({{b_2}}^{2}-{b_2}{b_4}+{{b_4}}^{2})]{s_2}\bigg){l}^{4}-2{l}^{2}{{z}}^{2}+6{s_2}{{z}}^{2}\bigg]{s_1}+\frac34{{z}}^{3}(-\frac13+\lambda)\bigg\}\frac{{{z}}^{2}{w}^{4}}{(3\lambda-1)({b_2}-{b_4})}=0,\nonumber
\eea
\bea
& &6{l}^{2}{s_1}{\kappa}^{4}\bigg\{\frac43({b_2}+{b_4}){{z}}^{3}+\bigg[\frac43
+\bigg(-5{{b_2}}^{2}+({b_1}-\frac{11}{3}{b_4}){b_2}+({b_1}-\frac{14}{3}{b_4}){b_4}\bigg){s_2}\bigg]{{z}}^{2}\nonumber\\
&+&{s_2}({b_1}-\frac{16}{3}{b_2}-5{b_4}){z}-{\frac{19}{3}}{s_2}\bigg\}+4\bigg\{\bigg[(-2{l}^{2}+6{s_2}){{z}}^{3}+[\lambda({b_2}+{
b_4}){l}^{2}-2{s_2}({b_1}+{b_2})]{l}^{2}{{z}}^{2}\nonumber\\
&+&{s_2}\bigg((9{{b_4}}^{2}+(-2{b_1}-{b_2}){b_4}+{b_2}(-5{b_2}+{b_1}))\lambda-\frac92{{b_4
}}^{2}+({b_1}+{b_2}){b_4}+\frac12{{b_2}}^{2}\bigg){l}^{4}{z}\nonumber\\
&-&4(({b_2}-2{b_4})\lambda+{b_4}){s_2}{l}^{4}\bigg]{s_1}+\frac34(3\lambda-1){{z}}^{4}\bigg\}\frac{{{z}}^{2}{w}^{4}}{(3\lambda-1)({b_2}-{b_4})}=0,\nonumber
\eea


\begin{thebibliography}{99}


\bibitem{Horava:2009uw}
  P.~Horava,
  Phys.\ Rev.\  D {\bf 79}, 084008 (2009)
  [arXiv:0901.3775 [hep-th]].



\bibitem{Lu:2009em}
  H.~Lu, J.~Mei and C.~N.~Pope,
  Phys.\ Rev.\ Lett.\  {\bf 103}, 091301 (2009)
  arXiv:0904.1595 [hep-th].

\bibitem{Horava:2008ih}
  P.~Horava,
  JHEP {\bf 0903}, 020 (2009)
  [arXiv:0812.4287 [hep-th]].


\bibitem{Horava:2009if}
  P.~Horava,
  Phys.\ Rev.\ Lett.\  {\bf 102}, 161301 (2009)
  arXiv:0902.3657 [hep-th].


\bibitem{Takahashi:2009wc}
  T.~Takahashi and J.~Soda,
  Phys.\ Rev.\ Lett.\  {\bf 102}, 231301 (2009)
  arXiv:0904.0554 [hep-th].

\bibitem{Calcagni:2009ar}
  G.~Calcagni,
  JHEP {\bf 0909}, 112 (2009)
  arXiv:0904.0829 [hep-th].

\bibitem{Kiritsis:2009sh}
  E.~Kiritsis and G.~Kofinas,
  Nucl.\ Phys.\  B {\bf 821}, 467 (2009)
  arXiv:0904.1334 [hep-th].

\bibitem{Kluson:2009sm}
  J.~Kluson,
  JHEP {\bf 0907}, 079 (2009)
  arXiv:0904.1343 [hep-th].


\bibitem{Mukohyama:2009gg}
  S.~Mukohyama,
  JCAP {\bf 0906}, 001 (2009)
  arXiv:0904.2190 [hep-th].

\bibitem{Nikolic:2009hr}
  H.~Nikolic,
  Int.\ J.\ Mod.\ Phys.\  A {\bf 25}, 1477 (2010)
  arXiv:0904.2287 [hep-th].

\bibitem{Brandenberger:2009yt}
  R.~Brandenberger,
  Phys.\ Rev.\  D {\bf 80}, 043516 (2009)
  arXiv:0904.2835 [hep-th].

\bibitem{Nikolic:2009jg}
  H.~Nikolic,
Mod.\ Phys.\ Lett.\  A {\bf 25}, 1595 (2010)
  arXiv:0904.3412 [hep-th].

\bibitem{Izawa:2009ne}
  K.~I.~Izawa,
  arXiv:0904.3593 [hep-th].

\bibitem{Nastase:2009nk}
  H.~Nastase,
  arXiv:0904.3604 [hep-th].

\bibitem{Pal:2009np}
  S.~S.~Pal,
  Class.\ Quant.\ Grav.\  {\bf 26}, 245014 (2009)
  arXiv:0904.3620 [hep-th].

\bibitem{Cai:2009pe}
  R.~G.~Cai, L.~M.~Cao and N.~Ohta,
  Phys.\ Rev.\  D {\bf 80}, 024003 (2009)
  arXiv:0904.3670 [hep-th].

\bibitem{Cai:2009ar}
  R.~G.~Cai, Y.~Liu and Y.~W.~Sun,
  JHEP {\bf 0906}, 010 (2009)
  arXiv:0904.4104 [hep-th].

\bibitem{Volovik:2009av}
  G.~E.~Volovik,
  JETP Lett.\  {\bf 89}, 525 (2009)
  arXiv:0904.4113 [gr-qc].

\bibitem{Piao:2009ax}
  Y.~S.~Piao,
  Phys.\ Lett.\  B {\bf 681}, 1 (2009)
  arXiv:0904.4117 [hep-th].

\bibitem{Gao:2009bx}
  X.~Gao,
  arXiv:0904.4187 [hep-th].

\bibitem{Colgain:2009fe}
  E.~O.~Colgain and H.~Yavartanoo,
  JHEP {\bf 0908}, 021 (2009)
  arXiv:0904.4357 [hep-th].

\bibitem{Sotiriou:2009gy}
  T.~Sotiriou, M.~Visser and S.~Weinfurtner,
  Phys.\ Rev.\ Lett.\  {\bf 102}, 251601 (2009)
  arXiv:0904.4464 [hep-th].

\bibitem{Chen:2009ka}
  B.~Chen and Q.~G.~Huang,
  Phys.\ Lett.\  B {\bf 683}, 108 (2010)
  arXiv:0904.4565 [hep-th].

\bibitem{Mukohyama:2009zs}
  S.~Mukohyama, K.~Nakayama, F.~Takahashi and S.~Yokoyama,
  Phys.\ Lett.\  B {\bf 679}, 6 (2009)
  arXiv:0905.0055 [hep-th].

\bibitem{Myung:2009dc}
  Y.~S.~Myung and Y.~W.~Kim,
  Eur.\ Phys.\ J.\  C {\bf 68}, 265 (2010)
  arXiv:0905.0179 [hep-th].

\bibitem{Cai:2009dx}
  R.~G.~Cai, B.~Hu and H.~B.~Zhang,
  Phys.\ Rev.\  D {\bf 80}, 041501 (2009)
  arXiv:0905.0255 [hep-th].

\bibitem{Orlando:2009en}
  D.~Orlando and S.~Reffert,
  Class.\ Quant.\ Grav.\  {\bf 26}, 155021 (2009)
  arXiv:0905.0301 [hep-th].

\bibitem{Gao:2009er}
  C.~Gao,
  Phys.\ Lett.\  B {\bf 684}, 85 (2010)
  arXiv:0905.0310 [astro-ph.CO].

\bibitem{Ha:2009ft}
  T.~Ha, Y.~Huang, Q.~Ma, K.~D.~Pechan, T.~J.~Renner, Z.~Wu and A.~Wang,
  arXiv:0905.0396 [physics.pop-ph].

\bibitem{Nishioka:2009iq}
  T.~Nishioka,
 Class.\ Quant.\ Grav.\  {\bf 26}, 242001 (2009)
 arXiv:0905.0473 [hep-th].

\bibitem{Kehagias:2009is}
  A.~Kehagias and K.~Sfetsos,
  Class.\ Quant.\ Grav.\  {\bf 26}, 242001 (2009)
  arXiv:0905.0477 [hep-th].

\bibitem{Konoplya:2009ig}
  R.~A.~Konoplya,
  Phys.\ Lett.\  B {\bf 679}, 499 (2009)
  arXiv:0905.1523 [hep-th].

\bibitem{Ghodsi:2009zi}
  A.~Ghodsi and E.~Hatefi,
  Phys.\ Rev.\  D {\bf 81}, 044016 (2010)
  arXiv:0906.1237 [hep-th].

\bibitem{Arnowitt:1962hi}
  R.~L.~Arnowitt, S.~Deser and C.~W.~Misner,
  arXiv:gr-qc/0405109.


\bibitem{Caldarelli:1998hg}
  M.~M.~Caldarelli and D.~Klemm,
  Nucl.\ Phys.\  B {\bf 545}, 434 (1999)
  [arXiv:hep-th/9808097].

\bibitem{Astefanesei:2006dd}
  D.~Astefanesei, K.~Goldstein, R.~P.~Jena, A.~Sen and S.~P.~Trivedi,
  JHEP {\bf 0610}, 058 (2006)
  [arXiv:hep-th/0606244].
  
\end{thebibliography}
\end{document}